\documentclass[aps, prl, reprint, twocolumn, flushbottom, superscriptaddress, floatfix]{revtex4-1}

\usepackage{graphicx}
\usepackage{amssymb}
\usepackage{gensymb}
\usepackage[usenames,dvipsnames]{color}
\usepackage{colortbl}
\usepackage{MnSymbol}
\DeclareSymbolFont{symbolsC}{U}{txsyc}{m}{n}
\DeclareMathSymbol{\Diamondblack}{\mathord}{symbolsC}{95}
\DeclareMathSymbol{\medbullet}{\mathbin}{symbolsC}{8}

\definecolor{navblue}{RGB}{0,0,170}

\begin{document}

\title{Probing of optical near-fields by electron rescattering on the 1\,nm scale}

\author{Sebastian Thomas}
\thanks{These authors contributed equally to this work.}
\affiliation{Max-Planck-Institut f\"ur Quantenoptik, 85748 Garching bei M\"unchen, Germany}
\author{Michael Kr\"uger}
\thanks{These authors contributed equally to this work.}
\affiliation{Max-Planck-Institut f\"ur Quantenoptik, 85748 Garching bei M\"unchen, Germany}
\author{Michael F\"orster}
\affiliation{Max-Planck-Institut f\"ur Quantenoptik, 85748 Garching bei M\"unchen, Germany}
\author{Markus Schenk}
\affiliation{Max-Planck-Institut f\"ur Quantenoptik, 85748 Garching bei M\"unchen, Germany}
\author{Peter Hommelhoff}
\email{peter.hommelhoff@mpq.mpg.de}
\affiliation{Max-Planck-Institut f\"ur Quantenoptik, 85748 Garching bei M\"unchen, Germany}
\affiliation{Friedrich-Alexander-Universit\"at Erlangen-N\"urnberg, 91058 Erlangen, Germany}

\date{\today}

\begin{abstract}
We present a new method of measuring optical near-fields within ${\sim} 1\,\mathrm{nm}$ of a metal surface, based on rescattering of photoemitted electrons. With this method, we  precisely measure the field enhancement factor for tungsten and gold nanotips as a function of tip radius. The agreement with Maxwell simulations is very good. Further simulations yield a field enhancement map for all materials, which shows that optical near-fields at nanotips are governed by a geometric effect under most conditions, while plasmon resonances play only a minor role. Last, we consider the implications of our results on quantum mechanical effects near the surface of nanostructures and discuss features of quantum plasmonics.
\end{abstract}

\maketitle

The excitation of enhanced optical near-fields at nanostructures allows the localization of electromagnetic energy on the nanoscale~\cite{Novotny2006a, Stockman2011}. At nanotips, this effect has enabled a variety of applications, most prominent amongst them are scanning near-field optical microscopy (SNOM)~\cite{Wessel1985, Inouye1994, Kawata2009, Schnell2011, Hartschuh2008}, which has reached a resolving power of $\mathrm{8\,nm}$~\cite{Raschke2005}, and tip-enhanced Raman spectroscopy (TERS)~\cite{Wessel1985, Stockle2000}. Because of the intrinsic nanometric length scale, measuring and simulating the tips' near-field has proven hard and led to considerably diverging results (see Refs.~\cite{Novotny2006a, Hartschuh2008} for overviews). Here we demonstrate a nanometric field sensor based on electron rescattering, a phenomenon well known from attosecond science~\cite{Corkum2007}. It allows measurement of optical near-fields, integrating over only $1\,\mathrm{nm}$ right at the structure surface, close to the length scale where quantum mechanical effects become relevant~\cite{Zuloaga2010, Marinica2012, Ciraci2012, Wachter2012, Teperik2013}. Hence, this method measures near-fields on a scale that is currently inaccessible to other techniques (such as SNOM or plasmonic methods in electron microscopy~\cite{GarciadeAbajo2010, Koh2011, Willets2012}), and reaches down to the minimum length scale where one can meaningfully speak about a classical field enhancement factor. In the future, the method will allow tomographic reconstruction of the optical near-field and potentially the sensing of fields in more complex geometries such as bow-tie or split-ring antennas.

\begin{figure}[t]
\includegraphics[width=\columnwidth]{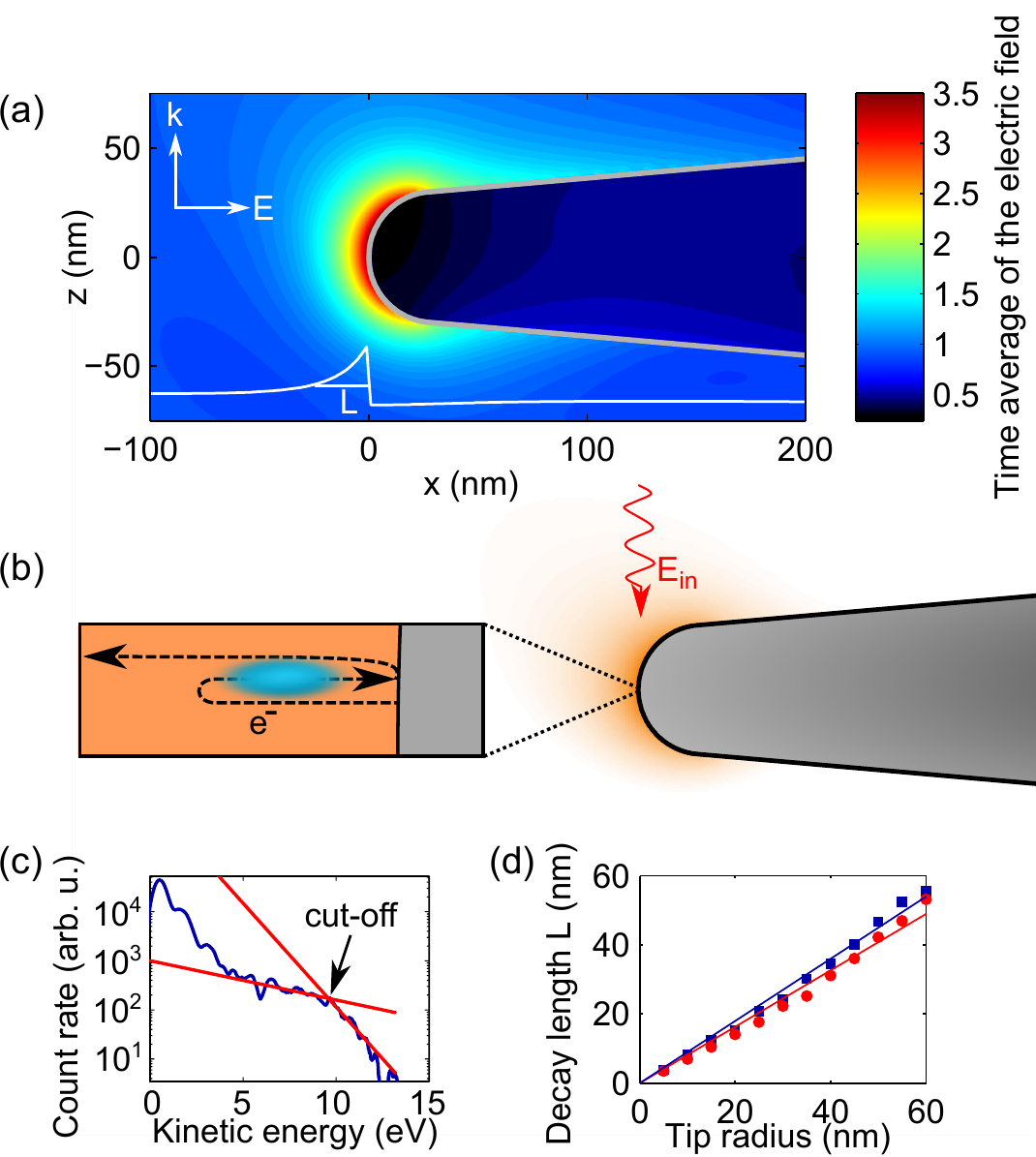}
\caption{(a) Time-averaged electric field (obtained from the simulation) near the apex of a tungsten tip ($R = 30\,\mathrm{nm}$, $\lambda = 800\,\mathrm{nm}$) in a plane spanned by the tip axis and the wave vector $\textbf{k}$ of the laser. The tip shape is indicated as a gray line. The white line at the bottom displays the near-field along $z = 0$ with the $1/e$ decay length $L$. The field rises from 1.2 to 3.4 over a distance of $29\,\mathrm{nm}$, where $1$ indicates the field strength in the bare laser focus without a tip. Note that the near-field is not symmetric with respect to the tip axis. This asymmetry is more prominent for larger tip radii~\cite{Yanagisawa2010}. (b) Illustration of electron rescattering: electrons are emitted in the optical near-field of a metal nanotip. A fraction of the emitted electrons is driven back to the tip surface, where they can scatter elastically. The kinetic energy gained during the rescattering process depends sensitively on the electric field near the tip surface. Thus the strength of the optical near-field is mapped to the kinetic energy of the emitted electrons. (c) Typical energy distribution of electrons emitted in the forward direction. The high-energy plateau ($\sim 5$ to $10\,\mathrm{eV}$) arises due to rescattering. Its cut-off is related to the local electric field amplitude at the metal. We obtain it from the intersection of two exponential fit functions (red lines). (d) Decay length $L$ as function of tip radius $R$ for tungsten tips ({\color{navblue} $\filledmedsquare$}) and gold tips ({\color{red} $\medbullet$}), deduced from simulations. The lines show linear fits: $L = (0.90 \pm 0.03) R$ for tungsten and $L = (0.82 \pm 0.04) R$ for gold. As the shape of the near-field mainly depends on the tip geometry, other materials behave very similarly.}
        \label{nfrescat}
\end{figure}

In general, three effects contribute to the enhancement of optical electric fields at structures that are smaller than the driving wavelength~\cite{Martin2001, Hartschuh2008, Zhang2009, Martin1997}. The first effect is geometric in nature, similar to the electrostatic lightning rod effect: the discontinuity of the electric field at the material boundary and the corresponding accumulation of surface charges lead to an enhanced near-field at any sharp protrusion or edge. This effect causes singularities in the electric field at ideal edges of perfect conductors. For real materials at optical frequencies, the electric field is not as strongly enhanced and remains finite~\cite{VanBladel1996}. The second effect occurs at structures whose size is an odd multiple of half the driving wavelength: optical antenna resonances can be observed there. The third effect concerns only plasmonic materials like gold and silver, where an enhanced electric field can arise due to a localized surface plasmon resonance. Antenna and plasmon resonances depend critically on the shape and material of the structure in question and occur only for specific wavelengths. In contrast, geometric effects are inherently broadband and result in a monotonically increasing field enhancement for increasing sharpness of the structure and for increasing discontinuity in the dielectric constant at the boundary. In spite of their different nature and properties, all three effects can be modeled in the framework of Maxwell's equations with linear optical materials. However, field enhancement calculations remain challenging because they crucially depend on the shape of the illuminated object, while analytic solutions of Maxwell's equations are known only for a few special cases like spheres and infinite cylinders. Accurate field enhancement measurements are equally challenging because of the nanometric length scale and the often unknown exact shape of the structure.

In this letter, we present experimental measurements with a new technique, the results of which we compare to numerical simulations of optical field enhancement at nanometric metal tips. Illuminating such a tip with light polarized parallel to the tip axis leads to the excitation of an enhanced near-field, which is spatially confined in all directions on the length scale of the tip radius~\cite{Martin2001, Zhang2009, Novotny2006a} (see Fig.~1(a)). The near-field drives a localized source of electrons at the tip apex~\cite{Hommelhoff2006, Hommelhoff2006a, Ropers2007}. Such photoemission experiments have found applications in a variety of different contexts aside from nanotips~\cite{Petek1997, Aeschlimann2007, Dombi2013}. Very recently, it has been observed that part of the electrons can be driven back to the parent tip within a single cycle of the optical field. There, the electrons can scatter elastically and gain more energy in the optical field~\cite{Kruger2011, Yalunin2011, Herink2012, Wachter2012}. This process, well known from atomic physics~\cite{Corkum1993, Corkum2007, Paulus1994a}, has been called rescattering and leads to pronounced spectral features that are sensitive to the local electric field. Here we exploit the rescattering effect to probe the near-field in the immediate vicinity of the tip surface, as illustrated in Fig.~\ref{nfrescat}(b).

Our experiment consists of an almost atomically smooth metal tip with a radius of curvature $R = 8$ to $50\,\mathrm{nm}$. Its apex lies in the focal spot of few-cycle laser pulses derived from a Ti:sapphire oscillator (wavelength $\lambda = 800\,\mathrm{nm}$, repetition rate $f_\mathrm{rep} = 80\,\mathrm{MHz}$, pulse duration $\tau \approx 6\,\mathrm{fs}$). The setup is described in more detail in Ref.~\cite{Schenk2010}. While this laser system reaches intensities of up to ${\sim}10^{12}\,\mathrm{W/cm}^2$ in the focus, we do not observe any influence of possible optical non-linearities on the field enhancement factor, and all our results (e.g., the linear dependence of the rescattering cut-off on laser intensity~\cite{Wachter2012}) are consistent with a linear model of the metal's optical response.

\begin{figure}[t]
\includegraphics[width=\columnwidth]{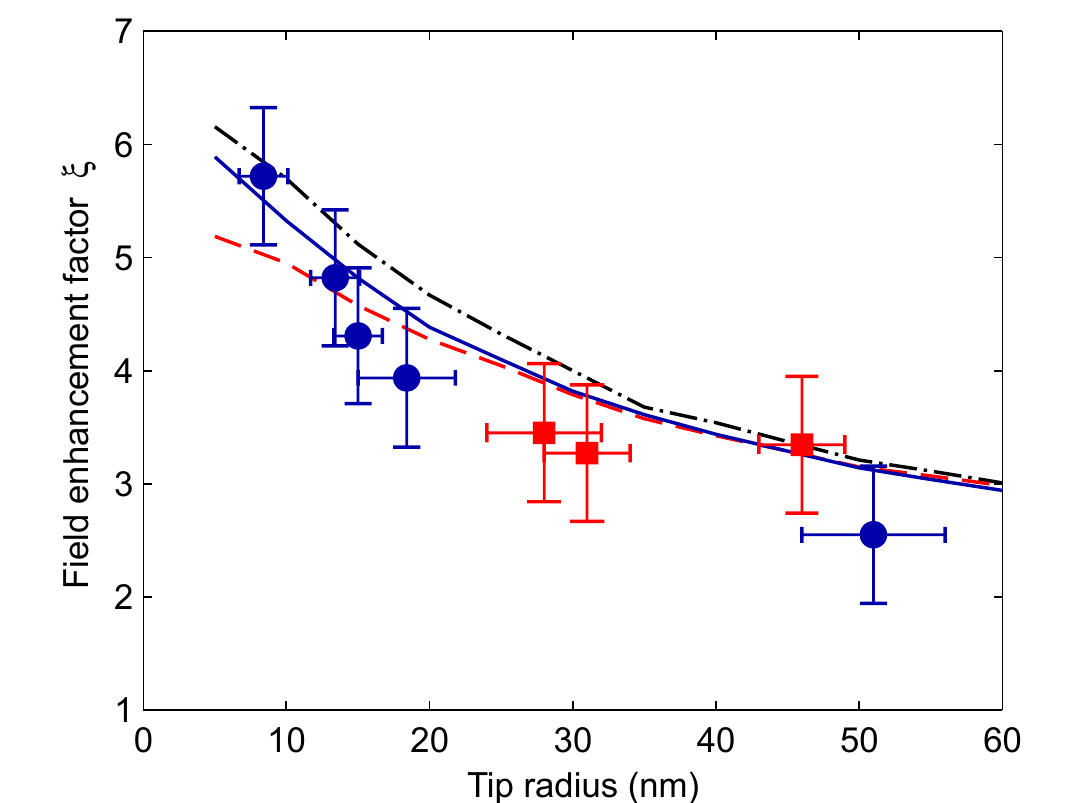}
\caption{Experimental results for the field enhancement factor of tungsten tips ({\color{navblue} $\medbullet$}) and gold tips ({\color{red} $\filledmedsquare$}) as a function of the tip radius. The uncertainty in $\xi$ represents an estimated systematic error due to the uncertainty in laser intensity. The lines are simulation results for $\lambda = 800\,\mathrm{nm}$ (W: solid blue line, Au: dashed red line, Ag: dash-dotted black line). The dielectric functions of the metals are taken from experimental data~\cite{Lide2004} (see Fig.~\ref{epsilon}(b)). For technical reasons related to mock surface plasmon reflection, we simulate gold and silver tips with a smaller opening angle than tungsten tips (W: $5\degree$, Au, Ag: $0\degree$). Simulations of tips with different angles show that this should not alter the results by more than $5\%$.}
    \label{radiusexp}
\end{figure}

Optical field enhancement enables us to observe electron rescattering at moderate pulse energies of less than $1\,\mathrm{nJ}$. We measure the energy distribution of the electrons emitted in the forward direction using a retarding field spectrometer. The recorded spectra yield information about the dynamics of the emitted electrons. A typical spectrum is shown in Fig.~\ref{nfrescat}(c). At small energies, such spectra display an exponential decrease in count rate, followed by a plateau towards larger energies. The latter is an indication of electron rescattering~\cite{Paulus1994a, Becker2002, Kruger2011, Kruger2012, Wachter2012}. This process has found utmost attention as it is at the core of attosecond science~\cite{Corkum2007}.

It has been shown that rescattering is highly sensitive to the peak electric field strength $E$ via the ponderomotive potential $U_{\mathrm{p}}$ the electron experiences in the light field~\cite{Paulus1994a, Becker2002, Kruger2012}: $U_{\mathrm{p}} = e^2  \lambda^2 E^2 / (16 \pi^2 m c^2).$ Here, $\lambda$ is the laser wavelength, $e$ and $m$ are the electron's charge and mass, and $c$ is the speed of light. The cut-off kinetic energy (see Fig.~\ref{nfrescat}(c)) after rescattering is given by $T_{\mathrm{cutoff}} = 10.007 \,U_{\mathrm{p}} + 0.538 \,\Phi$, where $\Phi$ denotes the tip's work function~\cite{Busuladzic2006}. Measuring $T_{\mathrm{cutoff}}$ hence yields $U_{\mathrm{p}}$~\cite{FESupp2012}.

Series of spectra for both tungsten and gold tips~\cite{Eisele2011} with various tip radii yield the dependence of the field enhancement factor on tip radius and material. We extract the cut-off position of the rescattering plateau and deduce, via the above expressions, the peak electric field $E$. We stress that $E$, the field acting on the electron, is the {\it enhanced} field present at the tip's surface. We thus obtain the field enhancement factor $\xi = E / E_{\mathrm{in}}$, with the laser field $E_{\mathrm{in}}$ deduced from intensity measurements. 

Figure~\ref{radiusexp} shows the field enhancement factor $\xi$ as a function of the tip radius $R$. For tungsten tips, $\xi$ grows by around a factor of $2$ with decreasing $R$, from $2.6 \pm 0.6$ at $(51 \pm 5)\,\mathrm{nm}$ to $5.7 \pm 0.6$ at $(8 \pm 2)\,\mathrm{nm}$. For gold nanotips with radii between $(46 \pm 3)\,\mathrm{nm}$ and $(28 \pm 4)\,\mathrm{nm}$, we obtain field enhancement factors between $\xi = 3.3 \pm 0.6$ and $3.5 \pm 0.6$. We have been unable to produce sharper gold tips with a well-controlled surface. Tip radii are determined in situ with the ring counting method in field ion microscopy or, for radii $> 20\,\mathrm{nm}$, using a scanning electron microscope~\cite{FESupp2012}.

We compare our results to fully independent simulations of field enhancement at tungsten, gold, and silver tips. They were performed using Lumerical (7.0.1), a commercial Maxwell solver implementing the finite-difference time-domain (FDTD) algorithm~\cite{Taflove2005}. From each simulation, we extract the field enhancement factor by fitting a quadratic decay to the near-field at the moment of greatest enhancement and extrapolating the result to the tip surface. This and other measures are essential to obtain meaningful results, as the finite mesh of the FDTD solver, together with the different length scales involved, makes this problem a tricky one. Further details and simulation results will be published elsewhere. The results for tungsten, gold, and silver are shown in Fig.~\ref{radiusexp}. Experimental and simulation results agree well within the error bars. Note that this agreement is obtained without any free parameters. Both experiment and simulation show that $\xi$ increases smoothly for sharper tips, an indication of field enhancement due to a geometric effect.

\begin{figure}[b]
\includegraphics[width=\columnwidth]{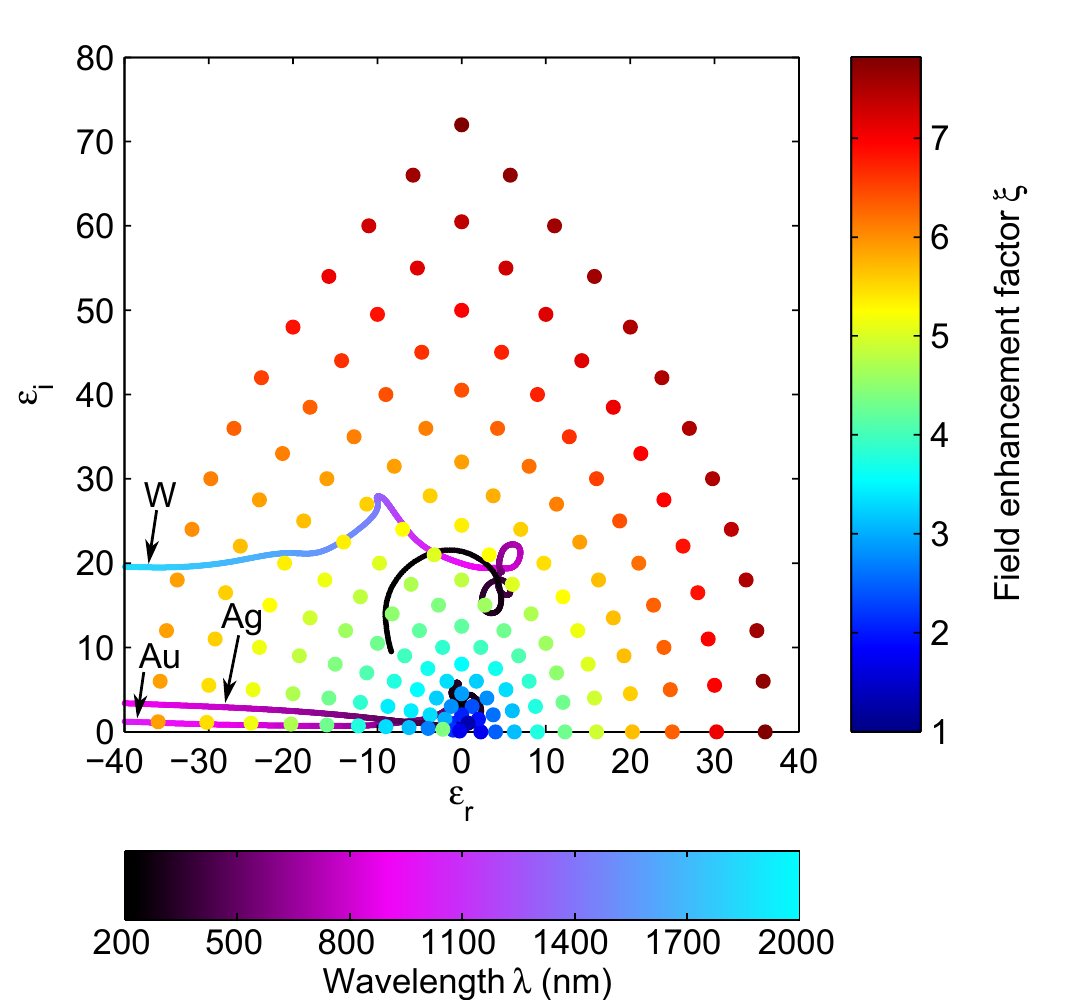}
\caption{Field enhancement factor as a function of the dielectric constant $\epsilon = \epsilon_\mathrm{r} + i \epsilon_\mathrm{i}$ obtained from simulations. The simulation parameters are $R = 10\,\mathrm{nm}$, $\lambda = 800\,\mathrm{nm}$, and an opening angle of $0\degree$. The dielectric constants of tungsten, gold, and silver at $800\,\mathrm{nm}$ are plotted in the complex plane for comparison (solid lines)~\cite{Haynes2011}. The right color scale applies to the dots, the bottom one to the lines. Note that the absolute value $|\epsilon|$ of the three materials is similar around $800\,\mathrm{nm}$.}
        \label{epsilon}
\end{figure}

Comparing our results to literature values of $\xi$, we find good agreement for tungsten tips (experiment~\cite{Neacsu2005, Yanagisawa2010}, theory~\cite{Martin2001, Yanagisawa2010}), while previous results for gold tips are inconsistent (experiment~\cite{Neacsu2005, Ropers2007}, theory~\cite{Martin2001, Bouhelier2003}) with some authors reporting much higher enhancement~\cite{Bouhelier2003, Neacsu2005, Ropers2007}. A possible explanation for this disagreement is that the near-field at plasmonic materials like gold is exceptionally sensitive to the geometry of the tip (the opening angle in particular~\cite{Martin2001, Zhang2009}) and its surface condition. This is supported by the large variance in enhancement factors at gold tips reported in Ref.~\cite{Neacsu2005}. Note also that far higher field enhancement factors are observed for tips in close vicinity ($ \lesssim R$) to surfaces~\cite{Yang2009}.

In our experiments with gold tips, we do not observe a large variance of field enhancement factors even though not all the tips had the ideal conical shape assumed in the simulations. A possible reason for this is that, before any measurement, we use field ion microscopy in conjunction with field evaporation to clean the tip surface and to ensure that the tip is almost ideally spherical in the vicinity of the apex~\cite{FESupp2012}. This is likely not the case in many other experiments. Evidently, more research is needed to fully understand the behavior of tips made of plasmonic materials. Such an investigation is beyond the scope of this letter. In the following analysis, we only consider perfectly smooth, conical tips (see Fig.~\ref{nfrescat}(a)) with small opening angles between $0\degree$ and $5\degree$. 

It appears, at first glance, surprising that the field enhancement factor of such different materials as tungsten and gold is so similar (see Fig.~\ref{radiusexp}), considering that gold supports the excitation of surface plasmons~\cite{Novotny2006a, Stockman2011}. We therefore analyze the dependence of the field enhancement factor on tip material in a series of simulations where we vary the complex dielectric constant $\epsilon = \epsilon_\mathrm{r} + i \epsilon_\mathrm{i}$ of the tip. This allows us to give a field enhancement map for all materials with $|\epsilon| \lesssim 40$, shown in Fig.~\ref{epsilon}. The results demonstrate that field enhancement occurs for any material with $\epsilon \neq 1$, even for pure dielectrics. Furthermore, the enhancement factor increases monotonically with the absolute value of the dielectric constant $|\epsilon|$, confirming that we observe field enhancement due to a geometric effect. We also note a slightly larger enhancement for $\epsilon_\mathrm{r} > 0$ than for $\epsilon_\mathrm{r} < 0$, which corresponds to an $\epsilon$-dependent phase shift ($< \pi$) of the near-field with respect to the driving field: the field enhancement factor is proportional to the maximum of the total electric field, which is reduced if the driving field and the near-field are out of phase. The $\epsilon$-dependent simulations reveal why the field enhancement factor of tungsten, gold, and silver tips is similar: they have a similar value of $|\epsilon|$ at $800\,\mathrm{nm}$.

In order to obtain higher enhancement factors, materials with larger values of $|\epsilon|$ are required. For example, we find $\xi = 7.6$ for $R = 10\,\mathrm{nm}$ aluminum tips ($\epsilon_\mathrm{Al} = -64 + 47i$ at $\lambda = 800\,\mathrm{nm}$, beyond the range of our simulations in Fig.~\ref{epsilon}). Alternatively, $\xi$ can be increased by using longer wavelengths, because both the tip sharpness relative to the wavelength and the absolute dielectric constant $|\epsilon|$ of many materials increase for longer wavelengths. We expect considerably higher field enhancement at sharp metal tips for mid- and far-infrared or terahertz radiation. An enhancement factor $\xi \approx 25$ has already been reported in SNOM experiments with terahertz radiation~\cite{Huber2008}.

In contrast to the increase with $|\epsilon|$, there is one point in Fig.~\ref{epsilon} close to $\epsilon = -2$ that shows a significantly higher enhancement than the points surrounding it. This can be interpreted as a localized plasmon resonance, similar to what is known from nanospheres~\cite{Martin2001}. It can be observed with a wavelength of $\lambda \approx 360\,\mathrm{nm}$ at silver tips~\cite{Zhang2009} or $\lambda \approx 520\,\mathrm{nm}$ at gold tips.

In the analysis of our experimental results, we have neglected the spatial variation of the near-field on the rescattered (field-probing) electrons' path, assuming instead a constant electric field. This is justified as the decay length of the near-field $L$ (see Fig.~\ref{nfrescat}(d)) is much longer than the maximum extension of the electron's path $M$: For our parameters, both classical~\cite{Kruger2011, Kruger2012} and quantum mechanical~\cite{Wachter2012, Yalunin2013} simulations indicate that the electrons' path extends approximately $1\,\mathrm{nm}$ from the surface before rescattering. On this scale, the sharpest tips we investigate show a near-field variation of ${\sim} 20\%$. Including this spatial variation into classical calculations of rescattering changes the enhancement factor by $0.4$ only, even for the sharpest tip in our experiments. This is less than the measurement uncertainty. For longer wavelengths or higher field strengths, the maximum extension $M$ increases. In this case, the effect may be more significant and can even suppress rescattering completely~\cite{Herink2012}.

One intriguing application of our method is the investigation of quantum effects in nanoplasmonics, a new field that has recently been named quantum plasmonics. Self-consistent quantum mechanical calculations of small nanoparticles (radius of curvature $< 2\,\mathrm{nm}$) illuminated by laser pulses show that the excited surface charge density, the root cause of the optical near-field, extends over several {\aa}ngstr{\" o}ms beyond the surface~\cite{Zuloaga2010}. This ``electron spill-out'' reduces the strength of the near-field by up to ${\sim} 50\%$. For small nanoparticles, it was shown that these effects are noticeable only at a distance of $Q < 0.5\,\mathrm{nm}$ from the surface, while the near-field retains its classical shape for larger distances. As fully quantum mechanical calculations of larger nanoparticles remain difficult (although large steps are being made in this direction~\cite{Teperik2013}), it is unclear if the length scale of nonclassical behavior $Q$ depends on the size of the nanoparticle. The authors of Ref.~\cite{Zuloaga2010} suspect $Q$ to increase for larger nanoparticles.

While a fully integrated quantum calculation of both field enhancement and electron rescattering is beyond the scope of this letter, we will discuss the effects of quantum plasmonics on rescattering qualitatively. They depend on three length scales: the extent of nonclassical field reduction $Q$, the near-field's decay length $L$, and the rescattered electron's maximum extension $M$. If $Q \approx L$ as in Ref.~\cite{Zuloaga2010}, the maximum of the near-field is significantly reduced, which implies a corresponding reduction of the cut-off energy. Extremely sharp nanostructures ($R \lesssim 3\,\mathrm{nm}$) will be required to reach this regime if $Q$ does not scale with structure size. As discussed earlier, rescattering may be suppressed in this case, depending on the relation of $M$ and $L$~\cite{Herink2012}.

If $L \gg Q$, only a small fraction of the near-field's extent is reduced in strength so that the maximum of the near-field is almost unchanged. In this case, quantum effects are only noticeable if $M \le Q$, because the rescattered electron would not be sensitive to the reduced field strength otherwise. The parameters in our experiments are $M \approx 1\,\mathrm{nm}$ (including a non-zero tunneling distance~\cite{Busuladzic2006, Kruger2012, Hickstein2012}) and $L \ge 8\,\mathrm{nm}$, so quantum effects should only be visible if $Q$ becomes larger than $0.5\,\mathrm{nm}$ for larger nanostructures. The agreement between experimental results and classical theory seems to suggest that $M > Q$. Hence, Q does not seem to scale with structure size, as hypothesized in Ref.~\cite{Zuloaga2010}. However, there is still the possibility of quantum plasmonic effects on a larger scale within the error bars of our results. An increase of $Q$ for larger tips might explain the steeper decrease of $\xi$ for larger radii we observe in the experiment as compared to the simulation (see Fig.~\ref{radiusexp}).

In conclusion, we demonstrate a new method of probing optical near-fields within $1\,\mathrm{nm}$ distance from the surface of a nanoscale metal tip. The method is based on rescattering of electrons driven by short laser pulses. The length scale on which the near-field is measured reaches down to dimensions that are of utmost interest in the emerging field of quantum plasmonics. Experimental results for the field enhancement factors of tungsten and gold tips agree well with Maxwell simulations. Based on these results, we give a field enhancement map for a wide range of materials. Furthermore, the simulations reveal that geometric effects are the predominant mechanism of optical field enhancement at nanotips in most cases. Exceptions exist close to plasmon resonances. In the future, a tomographic reconstruction of the near-field, likely in three dimensions, will be possible by measuring the cut-off energy of the rescattered electrons while varying the laser power or wavelength.

\begin{acknowledgments}
We would like to thank Peter Nordlander, Markus Raschke, and Hirofumi Yanagisawa for insightful discussions as well as Philipp Altpeter, Jakob Hammer, and Sebastian Stapfner for assistance with scanning electron microscope imaging.
\end{acknowledgments}

\end{document}


\title{Supplemental information to\\Probing of optical near-fields by electron rescattering on the 1\,nm scale}

\author{Sebastian Thomas}
\thanks{These authors contributed equally to this work.}
\affiliation{Max-Planck-Institut f\"ur Quantenoptik, 85748 Garching bei M\"unchen, Germany}
\author{Michael Kr\"uger}
\thanks{These authors contributed equally to this work.}
\affiliation{Max-Planck-Institut f\"ur Quantenoptik, 85748 Garching bei M\"unchen, Germany}
\author{Michael F\"orster}
\affiliation{Max-Planck-Institut f\"ur Quantenoptik, 85748 Garching bei M\"unchen, Germany}
\author{Markus Schenk}
\affiliation{Max-Planck-Institut f\"ur Quantenoptik, 85748 Garching bei M\"unchen, Germany}
\author{Peter Hommelhoff}
\email{peter.hommelhoff@mpq.mpg.de}
\affiliation{Max-Planck-Institut f\"ur Quantenoptik, 85748 Garching bei M\"unchen, Germany}
\affiliation{Friedrich-Alexander-Universit\"at Erlangen-N\"urnberg, 91058 Erlangen, Germany}

\date{\today}

\maketitle

\section{Photoemission and rescattering}
Here we give a short introduction to photoemission and rescattering from nanotips. For a much more detailed explanation, please see Ref.~\cite[S][]{Kruger2012}.

Based on Keldysh theory~\cite[S][]{Keldysh1965}, two regimes are usually distinguished in photoemission depending on the laser and material parameters: the multiphoton regime and the tunneling regime. In the first case, electron emission is described perturbatively as multiphoton absorption. In the second case, the surface barrier is bent so strongly that electrons can be emitted by tunneling. These two pictures represent limiting cases of a more general theory. They are distinguished by the Keldysh parameter
\begin{equation}
    \gamma = \sqrt{\frac{\phi}{2 U_\mathrm{p}}},
    \label{eq:keld}
\end{equation}
where $\phi$ denotes the material's work function and $U_\mathrm{p}$ the ponderomtive energy of the laser field. The multiphoton regime corresponds to $\gamma \gg 1$ while the tunneling regime corresponds to $\gamma \ll 1$. Our experiments are in a transition regime ($\gamma \approx 1$) where features of both pictures can be observed.

Our measurement of the near-field strength is not based on the photoemission itself, but on the electron's movement inside the laser field after the emission. As explained in the main text, some of the emitted electrons may be driven back to the tip surface and scatter elastically there, gaining more energy in this process. This rescattering process leads to the formation of a characteristic plateau in the emitted electrons' energy spectrum (shown in Fig.~1(c)). The plateau's cut-off energy, i.e., the highest energy the rescattered electrons can reach, is directly related to the electric field intensity that drives the rescattering process.

\section{Scaling of the rescattering cut-off energy with intensity}
The cut-off energy $T_\mathrm{cutoff}$ of rescattered electrons from atoms and molecules is known to follow the ``$10\,U_\mathrm{p}$ scaling law'' for Keldysh parameters $\gamma < 1$ according to classical models~\cite[S][]{Corkum1993,Paulus1994}. However, quantum corrections have to be considered since the influence of the binding energy (or work function) $\Phi$ is not negligibly small when $\gamma$ is increased~\cite[S][]{Busuladzic2006}. This leads to the approximate formula
\begin{equation}
T_{\mathrm{cutoff}} = 10.007\,U_\mathrm{p} + 0.538\,\Phi.
\label{eq:cut-off}
\end{equation}
This expression is a good approximation only for $\gamma < 1$ whereas the parameters of the enhanced near-field in our experiment result in $\gamma \approx 2$. We have therefore calculated the cut-off energy from quantum orbit theory~\cite[S][]{Salieres2001, Becker2002} and found reasonable agreement with Eq.~\ref{eq:cut-off} (deviation of less than ${12\%}$) for our parameters. For simplicity, we chose to follow Eq.~\ref{eq:cut-off} in the analysis of our experimental data.

In a further experimental and theoretical study~\cite[S][]{Wachter2012}, a field enhancement factor of $\xi \approx 7.5$ was determined for a tungsten tip with radius $R \approx 6\,$nm with a time-dependent density functional theory (TDDFT) simulation. Evaluating the experimental data with Eq.~\ref{eq:cut-off} gives a value of $\xi \approx 6.3$. Both values are in reasonable agreement with the measurement and simulation results for the field enhancement factors of this study. Simulating a $R = 5\,\mathrm{nm}$ tungsten tip yields $\xi = 5.9$. We conclude that the rescattering-based method of extracting $\xi$ is viable.

\section{Measurement of tip radii}
\begin{figure}[t]
\includegraphics[width=1.0\columnwidth]{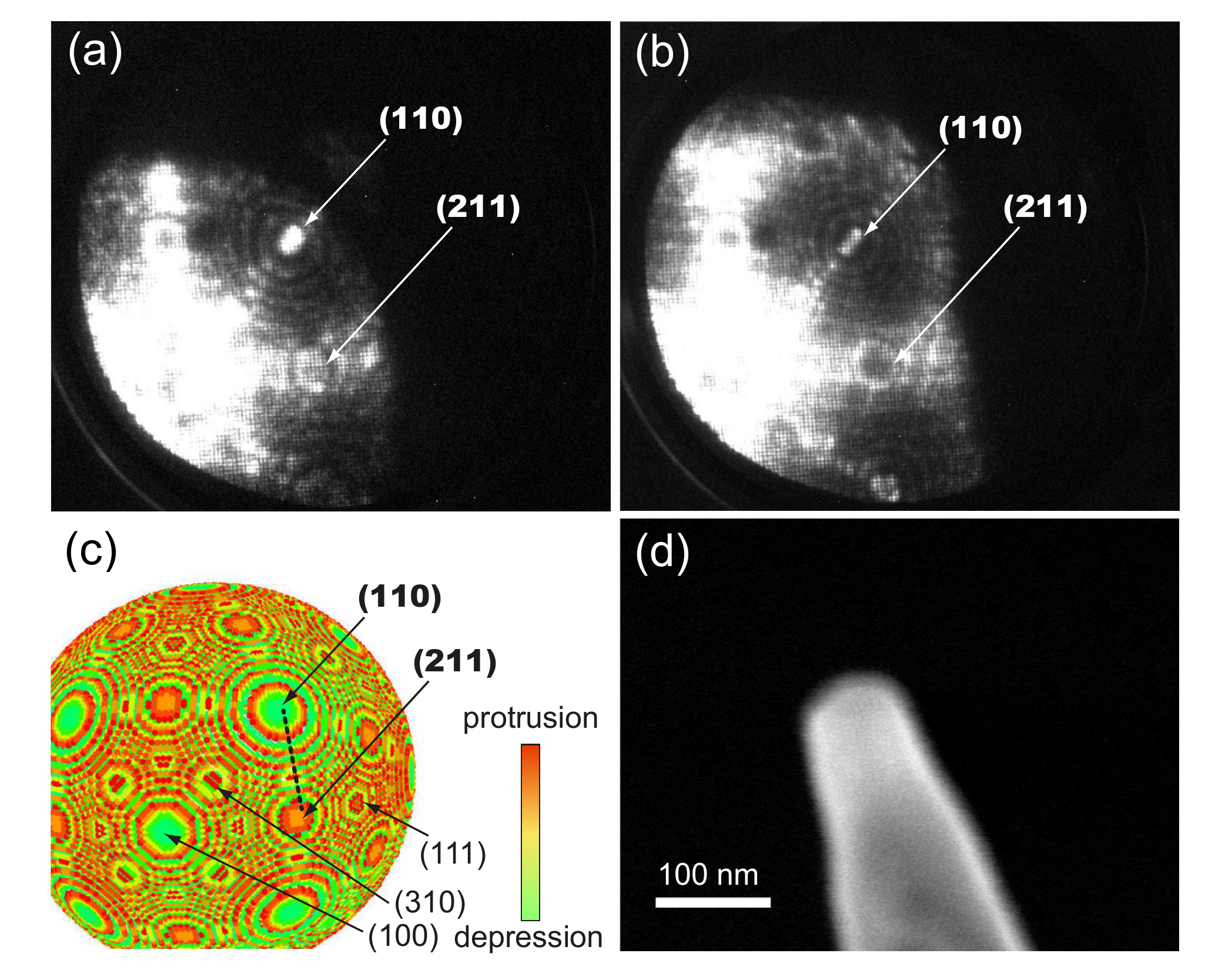}
\caption{Determination of tip radii by field ion microscopy and scanning electron microscope imaging. (a) Field ion microscope (FIM) image for the tungsten tip with a smaller radius of curvature. Counting the number of rings $n$ in the image between two crystallographic poles (in this case (110) and (211)) gives an estimate of the tip radius of curvature $R$ (here $n = 8\pm 1$, or $R = (13.4\pm 1.7)$\,nm). (b) The same for the tip with a slightly larger radius (here $n = 9\pm 1$, or $R = (15.0\pm 1.7)$\,nm). (c) Ball model of a tungsten tip in (310) orientation. The color map indicates the deviation of the surface atoms from an ideal hemisphere. Protruding atoms are visible as bright spots in FIM images. The ring counting method is applied between the (110) and (211) poles (dotted line). (d) Scanning electron microscope image of a gold tip ($R = (46 \pm 3)$\,nm).}
\label{fimtipradius}
\end{figure}

We produce tungsten and gold tips by electrochemical etching. To measure the field enhancement factor as a function of tip radius (Fig.~2), the tungsten tip is gradually blunted by field evaporation~\cite[S][]{Tsong1990} and, in the final step, by heating the tip slightly above $1000$\,K. Electron spectra are recorded for each blunting step in order to determine the cut-off energy for a given intensity (see Fig.~1(c) for an example). The tip radius of curvature $R$ is determined \emph{in situ} with the help of field ion microscopy~\cite[S][]{Tsong1990} (FIM) for tips with $R < 30$\,nm (see Fig.~\ref{fimtipradius}(a) and (b) for examples). The ring counting method~\cite[S][]{Tsong1990} gives a reliable estimate of the local radius of curvature of the surface: atoms terminating atomic layers of the tungsten bcc lattice structure at the surface are visible as bright spots in FIM because they protrude from the rest of the surface (see ball model in Fig.~\ref{fimtipradius}(c)). Around the (110) poles, pronounced ring structures are found. Counting the number of rings $n$ between two crystallographic poles gives the radius of curvature via the relation
\begin{equation}
r = \frac{n s}{1 - \cos \alpha},
\end{equation}
where $s = {a}/{(\delta\sqrt{h^{2}+k^{2}+l^{2}})}$ is the lattice step size for the reference orientation $(h,k,l)$ (here (110)). $a$ denotes the lattice constant (tungsten: $a = 3.16\,\mathrm{\AA}$) and $\delta$ is 1 if $h+k+l$ is an even number and 2 otherwise. $\alpha$ is the angle between reference orientation $(h,k,l)$ and secondary orientation $(h',k',l')$ (here (211)) and is given by $\cos\alpha=(hh'+kk'+ll')/\sqrt{(h^{2}+k^{2}+l^{2}) (h'^{2}+k'^{2}+l'^{2})}$. For $(h,k,l) = (110)$ and $(h',k',l') = (211)$, we find $\alpha = 30\degree$. After the final blunting step ($R > 30\,$nm), FIM is not possible due to high-voltage breakdown limitations. Therefore, we image the tip with a scanning electron microscope (SEM) after the measurements.

FIM of Au tips at room temperature cannot provide atomic resolution~\cite[S][]{Eisele2011}. Therefore, it is not possible to employ the ring counting method and we determine the tip radius from SEM imaging (see Fig.~\ref{fimtipradius}(d)).